
\input harvmac
%
%
\def\RF#1#2{\if*#1\ref#1{#2.}\else#1\fi}
\def\NRF#1#2{\if*#1\nref#1{#2.}\fi}
\def\refdef#1#2#3{\def#1{*}\def#2{#3}}
%
%
\def \ts{\thinspace}
\def \AJM{{\it Am.\ts J.\ts Math.\ts}}
\def \CMP{{\it Comm.\ts Math.\ts Phys.\ts }}
\def \NP{{\it Nucl.\ts Phys.\ts }}
\def \PL{{\it Phys.\ts Lett.\ts }}

\def \Tahoe{Proceedings of the XVIII
 International Conference on Differential Geometric Methods in Theoretical
 Physics: Physics and Geometry, Lake Tahoe, USA 2-8 July 1989}
\def \Tahoe{Proceedings of the NATO
 Conference on Differential Geometric Methods in Theoretical
 Physics, Lake Tahoe, USA 2-8 July 1989 (Plenum 1990)}
\def \Zm{Zamolodchikov}
\def \AZm{A.\ts B.\ts \Zm}
\def \AlZm{Al.\ts B.\ts \Zm}
\def\me{P.\ts E.\ts Dorey}
\def\dur{H.\ts W.\ts Braden, E.\ts Corrigan, \me\ and R.\ts Sasaki}
%
%
\refdef\rAFZa\AFZa{A.\ts E.\ts Arinshtein, V.\ts A.\ts Fateev and
 \AZm, \PL {\bf B87} (1979) 389}
\refdef\rBa\Ba{H.\ts W.\ts Braden, \lq A note on affine Toda couplings',
 Edinburgh preprint 91-01}
\refdef\rBc\Bc{N.\ts Bourbaki, {\it Groupes et alg\`ebres de Lie} {\bf
 IV, V, VI,} (Hermann, Paris 1968)}
\refdef\rBCDSa\BCDSa{\dur, \PL {\bf B227} (1989) 411}
\refdef\rBCDSb\BCDSb{\dur, \Tahoe}
\refdef\rBCDSc\BCDSc{\dur, \NP {\bf B338} (1990) 689}
\refdef\rBCDSd\BCDSd{\dur,
 ``Aspects of affine Toda field theory", to appear in the Proceedings
of the $10^{\rm th}$ Winter School on Geometry and Physics, Srni,
Czechoslovakia; Integrable Systems and Quantum Groups, Pavia, Italy;
Spring Workshop on Quantum Groups, ANL, USA}
\refdef\rBCDSe\BCDSe{\dur, \NP {\bf B356} (1991) 469}
\refdef\rBLa\BLa{D.\ts Bernard and A.\ts LeClair, \lq\lq Quantum group
 symmetries and non-local currents in 2D QFT", preprint CLNS-90/1027,
 SPhT-90/144, \CMP\ to appear}
\refdef\rBPZa\BPZa{A. A. Belavin, A. M. Polyakov and A. B. Zamolodchikov, \NP
 {\bf B241} (1984) 333}
\refdef\rBSa\BSa{H.\ts W.\ts Braden and R.\ts Sasaki, \PL {\bf B255} (1991)
343}
\refdef\rCa\Ca{P. Christe, ``S-matrices of the tri-critical
 Ising model and Toda systems", \Tahoe}
\refdef\rCf\Cf{R.\ts Carter, {\it Simple Groups of Lie Type}, (Wiley
 1972)}
\refdef\rCg\Cg{H.\ts Coxeter, {\it Am. J. Math.} {\bf 62} (1940) 457}
\refdef\rCh\Ch{H.\ts Coxeter, {\it Regular Polytopes}, (Methuen 1948)}
\refdef\rCi\Ci{\Cg\semi\Ch}
\refdef\rCMa\CMa{P.\ts Christe and G.\ts Mussardo, {\it Nucl. Phys}.
 {\bf B330} (1990) 465}
\refdef\rCMb\CMb{P.\ts Christe and G.\ts Mussardo,
 {\it Int.~J.~Mod.~Phys.}~{\bf A5} (1990) 4581}
\refdef\rCMc\CMc{J.\ts Cardy and G.\ts Mussardo, \PL {\bf B225} (1989) 275}
\refdef\rDc\Dc{\me,  \NP {\bf B358} (1991) 654}
\refdef\rDDa\DDa{C.\ts Destri and H.\ts J.\ts de Vega, {\it Phys. Lett.}
 {\bf B233} (1989) 336}
\refdef\rFb\Fb{M.\ts D.\ts Freeman, \PL {\bf B261} (1991) 57}
\refdef\rFJa\FJa{I.B. Frenkel and N. Jing, {\it Proc. Nat. Acad. Sci.
U.S.A.} {\bf 85} (1988) 9373}
\refdef\rFKb\FKb{I. B. Frenkel and V. G. Kac, {\it Inv. Math.} {\bf 62} (1980)
 23}
\refdef\rFKMa\FKMa{P.\ts G.\ts O.\ts Freund, T.\ts Klassen and E.\ts Melzer,
 {\it Phys. Lett.} {\bf B229} (1989) 243}
\refdef\rFLOa\FLOa{A.\ts Fring, H.\ts C.\ts Liao and D.\ts Olive,
  \lq The mass spectrum and coupling in affine Toda theories'
   preprint IC/TP-90-91/25}
\refdef\rFZa\FZa{V.\ts A.\ts Fateev and \AZm, {\it Int. J. Mod. Phys.} {\bf A5}
 (1990) 1025}
\refdef\rGa\Ga{C.\ts J.\ts Goebel, {\it Prog. Theor. Phys. Suppl.} {\bf 86}
 (1986) 261}
\refdef\rHc\Hc{T.J. Hollowood, private communication}
\refdef\rHMa\HMa{T.\ts J.\ts Hollowood and P.\ts Mansfield, \PL {\bf B226}
 (1989) 73}
\refdef\rJb\Jb{M.\ts Jimbo, {\it Int. J. Mod. Phys.} {\bf A4} (1989)
 3759, in {\it Braid Group, Knot Theory and Statistical Mechanics} (World
 Scientific 1989)}
\refdef\rKa\Ka{M.\ts Karowski, \NP {\bf B153} (1979) 244}
\refdef\rKb\Kb{B.\ts Kostant, \AJM {\bf 81} (1959) 973}
\refdef\rKSa\KSa{R. K\"oberle and J. A. Swieca, \PL {\bf B86}
 (1979) 209}
\refdef\rKTa\KTa{M. Karowski and H. Thun, {\it Nucl. Phys.} {\bf B130}
 (1977) 295}
\refdef\rKMa\KMa{T.\ts R.\ts Klassen and E.\ts Melzer, {\it Nucl. Phys.}
 {\bf B338} (1990) 485}
\refdef\rLWb\LWb{W.\ts Lerche and N.\ts P.\ts Warner, \NP {\bf B358}
(1991) 571}
\refdef\rMb\Mb{G. Mussardo, ``Away from criticality: some results from
 the S-matrix approach", \Tahoe}
\refdef\rMCa\MCa{\Ca\semi\Mb\semi\CMb}
\refdef\rMd\Md{N.\ts J.\ts MacKay, \NP {\bf B356} (1991) 729}
\refdef\rMJa\MJa{M. Jimbo and T. Miwa, Vertex Operators in Mathematics and
Physics, eds J. Lepowsky, S. Mandelstam and I.M. Singer, MSRI {\bf\# }3
(Springer-Verlag New York 1985)}
\refdef\rMOPa\MOPa{A. V. Mikhailov, M. A. Olshanetsky and A. M. Perelomov, {\it
 Comm. Math. Phys.} {\bf 79} (1981) 473}
\refdef\rOTa\OTa{D. I. Olive and N. Turok, {\it Nucl. Phys.} {\bf B215} (1983)
 470}
\refdef\rOTb\OTb{D. I. Olive and N. Turok, {\it Nucl. Phys.} {\bf B265}
(1986) 469}
\refdef\rOTc\OTc{D. I. Olive and N. Turok, unpublished manuscript}
\refdef\rOWa\OWa{E.\ts Ogievetsky and P.\ts Wiegmann,
   {\it Phys. Lett.} {\bf B168} (1986) 360}
\refdef\rPc\Pc{V.\ts Pasquier, {\it Nucl. Phys.} {\bf B285} (1987) 162}
\refdef\rPe\Pe{A. M. Polyakov, {\it JETP Letters} {\bf 12} (1970) 381}
\refdef\rYGa\YGa{See for example several of the articles in: C. N. Yang and M.
 L. Ge, {\it Braid Group, Knot theory and Statistical mechanics,} (World
 Scientific 1989)}
\refdef\rZa\Za{\AZm, ``Integrable Field Theory from Conformal Field Theory",
 Proceedings of the Taniguchi Symposium, Kyoto (1988)}
\refdef\rZb\Zb{\AZm, {\it Int. J. Mod. Phys.} {\bf A4} (1989) 4235}
\refdef\rZc\Zc{\AZm, {\it JETP Letters} {\bf 43} (1986) 730}
\refdef\rZd\Zd{\AZm, {\it Int. J. Mod. Phys.} {\bf A3} (1988) 743}
\refdef\rZe\Ze{\AZm, {\it Sov. Sci. Rev., Physics}, {\bf v.2} (1980)}
\refdef\rZf\Zf{\AZm, {\it Teor. Mat. Fiz.} {\bf 65} (1985) 347}
\refdef\rZh\Zh{\AZm, ``Exact Solutions of Conformal Field Theory in Two
Dimensions and Critical Phenomena", Kiev IMP preprint 87-65P (1987)}
\refdef\rZt\Zt{\AZm, Talk given in Oxford, January 1989}
\refdef\rZz\Zz{\Za\semi\Zb}
\refdef\rZg\Zg{\AlZm, ``Thermodynamic Bethe Ansatz in Relativistic Models.
 Scaling 3-state Potts and Lee-Yang Models", Moscow ITEP preprint (1989)}
\refdef\rZZa\ZZa{\AZm\  and \AlZm, {\it Ann. Phys.}
 {\bf 120} (1979) 253}
\def\tilde{\widetilde}

\def\({\left(}
\def\){\right)}

\def\ln{{\rm ln}}

\def\CT{{\cal T}}

\def\B{\bullet} \def\W{\circ}
\def\BS{\{\B\}} \def\WS{\{\W\}}
\def\wb{w_{\BS}} \def\ww{w_{\WS}}
\def\tw{\tilde w}
\def\bl#1{\{#1\}_+}
\def\mbl#1{\{#1\}_-}
\def\sbl#1{\bigl(#1\bigr)_+}

\def\ubl#1{\{#1\}}
\def\usbl#1{\bigl(#1\bigr)}
\def\hprod{\prod^{h-1}_{p=0}}
\def\hsum{\sum^{h-1}_{p=0}}

 \def\R{{\bf R}}

\def\q{{\cal Q}} 
 
\def\a{\alpha} \def\b{\beta} \def\g{\gamma} 
\def\d{\delta}
\def\sg{\sigma}
\def\l{\lambda} \def\p{\phi} \def\t{\theta} \def\o{\omega}
\def\ai{\a_{(i)}} \def\aj{\a_{(j)}} \def\ak{\a_{(k)}}
 \def\ajo{\a^o_{(j)}} \def\ako{\a^o_{(k)}}

\def\G{\Gamma}
\def\ha{\hat\alpha}  
\def\I{{\bf I}}
 
\def\ts{\thinspace}

\def\strutdepth{\dp\strutbox} 
\def\probsymbol{\vtop to \strutdepth{\baselineskip\strutdepth
  \vss\llap{\bf ??~~}\null}}
\def\prob#1{\strut\vadjust{\kern-\strutdepth\probsymbol}{\bf #1}}
\def \clap#1{\hbox to 0pt{\hss #1\hss}}
\def \label#1{\lower2ex\clap{$\scriptstyle #1$}}
\def \tick#1.#2.{\vrule height#1pt depth#2pt}
\def \seg#1.#2.{{\hfil \tick#1.#2.}}
\newcount\n
\def \tline#1.#2.#3.{{\n=0 \loop\ifnum\n<#1 \advance\n by1 {\hfil
   \tick#2.#3.}\repeat}}
\def \tickline#1.#2.#3.#4.#5.#6.
 {\n=0 \vbox{\hrule \vskip-#3pt
   \hbox to #1in{\label{#5}\tick#3.#4.
   \loop\ifnum\n<#2 \advance\n by1 \seg#3.#4.\repeat\label{#6}}}}
\def \ha{\hskip .5pt}
\def \bx#1.{{\ha\vrule depth#1pt \hrulefill\hrulefill\vrule depth#1pt\ha}}
\def \nobx{{\ha\ha\hskip .8pt\hfill\hfill}}
\def \nobxh{{\ha\hskip .4pt\hfill}}
\def \bxes#1.#2.{\n=0\loop\ifnum\n<#2 \advance\n by1 \bx#1.\repeat}
\def \nobxes#1.{\n=0\loop\ifnum\n<#1 \advance\n by1 \nobx\repeat}
\def \bag#1.#2.{\vtop {\hbox to #1{\hskip.2pt #2\hskip.2pt }}}
\def \bagh#1.#2.{\vtop {\hbox to #1{\hskip.2pt \nobxh #2\nobxh \hskip.2pt }}}

\Title{RIMS-787---SPhT-91/140 (corrected)
}{Root Systems and Purely Elastic S-Matrices II}
\centerline{
Patrick Dorey\footnote{$^\dagger$}{dorey@poseidon.saclay.cea.fr}}
\bigskip\centerline{Service de Physique Th\'eorique de
Saclay\rlap,\foot{{\it Laboratoire de la Direction des Sciences
de la Mati\`ere du Commissariat \`a l'Energie Atomique}}}
\centerline{91191 Gif-sur-Yvette cedex, France}
\vskip .5in

Starting from a recently-proposed general formula,
various properties of the ADE series of
purely elastic S-matrices
are rederived in a universal way. In particular,
the relationship between the pole structure and the
bootstrap equations is shown to follow from properties of root systems.
The discussion leads to a formula for the signs of the three-point
couplings in the simply-laced affine Toda theories, and a simple proof
of a result due to Klassen and Melzer of relevance to Thermodynamic
Bethe Ansatz calculations.

\Date{August 1991}

\NRF\rZz\Zz\NRF\rFZa\FZa
\NRF\rAFZa\AFZa
\NRF\rBCDSc\BCDSc
\NRF\rBCDSe\BCDSe\NRF\rBSa\BSa
\NRF\rCMa{\CMa\semi\CMb}
\NRF\rDDa\DDa\NRF\rKMa\KMa
\newsec{Introduction}
Occasionally, an integrable perturbation of a conformal field
theory results in a massive scattering theory which is purely
elastic, in that the S-matrix is diagonal in a suitable basis\ts\rZz.
This observation
has lead to some work on such S-matrices as interesting objects in
their own right. A series of examples connected with the ADE series
of Lie algebras has been uncovered, both directly in the context of
perturbed conformal field theory\ts\rFZa, and also via
the study of affine Toda field theories\ts\refs{\rAFZa {--}\rKMa}
(in fact, there are slight differences between the S-matrices found
in the two contexts; these will be mentioned where relevant).
As is often the case in the study of integrable quantum field
theories, the proposed S-matrices have not (at least so far) been
derived from first principles, but rather deduced on the basis of certain
assumptions and consistency requirements.
However, for a theory with a diagonal S-matrix, the Yang-Baxter
equation -- often a very powerful tool in the study of the
S-matrices of integrable theories\ts\RF\rZZa{See for example \ZZa}\
-- is trivially satisfied, and so gives no information. There does
remain the possibility that two particles in the theory may fuse to
form a third as a bound state\ts\RF\rKa\Ka. As emphasised by
Zamolodchikov\ts\rZz, for purely elastic scattering the resulting
bootstrap equations are sufficiently simple to provide a useful set of
consistency conditions, constraining both the conserved charges and
the S-matrix.

The nature of the bootstrap solutions for the ADE theories
is in fact closely linked to properties of the corresponding root
systems, and in particular the action on these root systems of the
Coxeter element of the Weyl group\ts\RF\rDc\Dc\def\I{\rDc}.
The aim of this paper
is to explore this connection a little further, with particular
emphasis on the implications for the structure of the S-matrix
elements.
By refining the notations used in \I, it turns out to be possible to
streamline the discussion considerably. After a description  of
some necessary formulae, section two outlines how this goes.
While this section contains no new results, it does give a proof of
the S-matrix bootstrap equations which simplifies and clarifies that
given previously.
Section three is devoted to a discussion of the pole structure of
S-matrix, and shows how this is exactly in accordance with the bound-state
structure predicted by the fusings. Various empirically observed
features of the purely elastic S-matrices turn out to be simple
consequences of the properties of root systems.
An application for some of the calculations in section three is given
in section four, giving a description of one set of signs for
the three-point couplings of
the simply-laced affine Toda theories, in terms of
roots and weights.
Finally, section five gives a
universal proof of an elegant formula due to Klassen
and Melzer\ts\RF\rKMa\KMa, and section six contains some concluding
remarks.

\newsec{Preliminaries}

Since this paper is a direct sequel to \I, the reader is referred
back to that paper for details of the motivations and many of the original
formulae. As mentioned in the introduction, elements of
the discussion given there become rather clearer if the notation
(in particular, the labelling of the orbits of the Coxeter element), is
changed slightly. For this, some results to be found in a
paper by Kostant\ts\RF\rKb\Kb\ will be needed, and this section starts
with a brief review of the relevant material.

First though, note that
various of the definitions and results to be given in this
section work equally well for simply-laced and non simply-laced root
systems. For example, the discussion in \rKb\ makes no distinction
between the two cases. It is also worth noting that the discussion
in \I\ of the conserved charge bootstrap goes through
essentially unchanged for the non simply-laced root systems.
However, the S-matrix formulae given in that paper seem to be hard to
generalise beyond the ADE series, possibly reflecting the difficulties
that were found in the (quantum) problem of finding
S-matrices for the non simply-laced affine Toda
theories\ts\refs{\rBCDSc,\rCMa}.
For this paper, then, attention will be restricted to the already-known
purely elastic scattering theories, that is to
those associated with the simply-laced Lie algebras. The discussion
will be based on a simply-laced root system
$\Phi$, of rank $r$, with
$\{\a_i\}$ a set of simple roots.
Letting $w_i$ denote the Weyl reflection corresponding to the simple
root $\a_i$ (so $w_i(x)=x-{2\over \a_i^2}(\a_i,x)\a_i$), set
$$w=w_1w_2\dots w_r,$$
so that $w$ is a Coxeter element. Also, let $\vev{w}$ be the subgroup
of $W$, the Weyl group, generated by $w$. Finally, for $i=1\dots r$ define
a root $\p_i$ by
\eqn\thetadf{
\phi_i=w_rw_{r-1}\dots w_{i+1}(\gamma_i).}
Then the following results are given in \rKb:\nobreak
\item{(i)} With the definition of positive and negative roots implied by the
given choice of simple roots, $\phi_i>0$ and $w(\phi_i)<0$.
\item{(ii)} If $\alpha$ is a root such that $\alpha>0$ and $w(\alpha)<0$, then
$\alpha$ is one of the  $\phi_i$.
\item{(iii)} Let $\G_i$ be the orbit of $\phi_i$ under $\vev{w}$. Then the
$\G_i$'s are disjoint, each has $h$ elements, and thus their union is
all of $\Phi$.

\noindent
The set $\{\phi_i\}$ possesses one further useful property, which
can be found for example
in \RF\rBc{\Bc: {\bf VI}, 1, exerc. 22}. If $\l_i$ is
the fundamental weight corresponding to the simple root $\a_i$, then
\eqn\bourbaki{\p_i = (1-w^{-1})\l_i. }
(To prove this result, note that
$w_i(\l_j)=\l_j-\delta_{ij}\a_j$,
which follows since the fundamental weights are dual
to the simple co-roots $\a_i^{\vee}\equiv {2\over\a_i^2} \a_i$.
For the simply-laced cases of interest here, $\a_i^2=2$ and
the $\l_i$ are dual to the simple roots themselves.)
This relation can be inverted, $w$ having no eigenvalue
equal to one. Writing
\eqn\rdef{(1-w^{-1})^{-1}=R,}
it is easily checked that
\eqn\rprop{R={1\over h}\sum^h_1 pw^p=-{1\over h}\hsum pw^{-p}.}
This mapping is not orthogonal, but rather satisfies
\eqn\rpropi{(R\a,\b)+(\a,R\b)=(\a,\b).}
Other identities, such as $(R\a,\b)=-(w\a,R\b)$, can also be
found but only \rpropi\ will be used below.
Projectors onto the various eigenspaces of $w$ are given by
\eqn\projop{P_s={1\over h}\hsum\o^{-sp}w^p={1\over h}\hsum\o^{sp}w^{-p},}
with corresponding eigenvalues $\o^s=e^{2\pi is/h}$. If $s$ is not an
exponent of the algebra,
then the spin $s$ eigenspace is null, and $P_s=0$.

It is often useful to focus on a particular
ordering of the simple roots, linked to a
two-colouring of the Dynkin diagram. This ordering is such that
$\{\a_i\}$ splits into two subsets, each of which
contains only mutually orthogonal roots:
\eqna\twosplit
$$ \{\a_1,\a_2,\dots\a_r\}
 = \{ \alpha_1,\dots \alpha_k \} \cup \{ \a_{k+1},\dots \a_r
 \}\eqno{\twosplit a}$$
Defining subsets of the indices $1,\dots r$ as
$$\B=\{1,2,\dots k\},\qquad \W=\{k+1,k+2,\dots r\}\eqno{\twosplit b}$$
(`black' and `white'),
the internal orthogonality amounts to the requirement that for $i\neq j$,
$(\a_i,\a_j)=0$ if $i$ and $j$ have the same colour.
By a small abuse of notation, the symbols
$\B,\B',\W,\W'$ and so on will occasionally be used to denote arbitrary
indices taken from the corresponding (black or white) subset.
A slightly different
notation was used in \I: black simple roots were called `type alpha',
white ones `type beta'.

In this ordering,
$$w=w_{\BS}w_{\WS}$$
with
$$w_{\BS}=\prod_{i\in\B}w_i,\qquad w_{\WS}=\prod_{j\in\W}w_j.$$
The internal orthogonality of the two subsets of the simple
roots implies that the reflections for simple roots of the same colour
commute, and that a reflection for a given simple root will leave
invariant all other simple roots of the same colour.
It follows that
\eqn\throot{\eqalign{\phi_{\B}&=w_{\WS}(\alpha_{\B}),\cr
\phi_{\W}&=\a_{\W}.\cr}}
In \I, the coset representatives were $\a_{\B}$
and $-\a_{\W}$; since
both $w_{\WS}$ and $-1$ induce charge
conjugation on the cosets, this means that, strictly speaking,
the assignment of cosets to simple roots induced by the $\phi_i$ differs by
an overall charge conjugation from that used in the earlier paper.
However this is merely a matter of labelling
convention -- for example,
one can interchange the values of the conserved charges on
particle and antiparticle simply by negating the normalisations of the
even spin charges -- and so can be ignored.

The ADE S-matrices will be built as products of
functional \lq building blocks'. In \I, the blocks used were
\eqn\block{\bl{x} =\cases{
{\displaystyle\sbl{x-1}\sbl{x+1}} &(perturbed conformal)\cr
\noalign{\vskip3pt}
{\displaystyle {\sbl{x-1}\sbl{x+1}\over \sbl{x-1+B}\sbl{x+1-B}}} &(affine
Toda)\cr}}
where
\eqn\sbblock{ \sbl{x}=\sinh\Bigl({\theta\over 2}+{i\pi x\over 2h}\Bigr)}
and
\eqn\bdep{B(\beta) = {1\over 2\pi}{\b^2\over 1+\b^2/4\pi}.}
The block for an affine Toda theory contains an extra coupling-constant
dependent
part over the `minimal' version suitable for perturbed conformal theories.
This turns out to have no effect on the physical pole structure ($\b$ being
real, and $B(\b)$ therefore
between $0$ and $2$), and so it is reasonable to use the same
notation for both cases.
In fact, the discussion of pole structure will be a little
more transparent if this block is swapped for another, namely
\eqn\mblock{\mbl{x}=\cases{
 -\bl{-x}^{-1} &(perturbed conformal)\cr
\noalign{\vskip2.5pt}
 \bl{-x}^{-1} &(affine Toda)\cr}}
Of course, any formula involving $\mbl{x}$ can be immediately rewritten in
terms of $\bl{x}$.

One further piece of notation will be needed. For any pair of roots
$\a,\b\in\Phi$, an integer $u(\a,\b)$ can be defined modulo $2h$ via the
following relations:
\eqn\udef{\eqalign{u(\a,\b)=-u(\b,\a)&\qquad u(w\a,\b)=u(\a,\b)+2\cr
u(\p_{\B},\p_{\B'})=u(\p_{\W},\p_{\W'})=0&\qquad u(\p_{\W},\p_{\B})=1.\cr}}
For the coset representatives $\p_i$, the abbreviated notation
$u_{ij}\equiv u(\p_i,\p_j)$ will often be used.
The definition is natural in that ${\pi s\over h}u(\a,\b)$ is the (signed)
angle between the projections of the roots $\a$ and $\b$ into the $\o^s$
eigenspace of $w$. Now the three-point couplings are described by the
following fusing rule\ts\I\ (also relevant in other contexts\ts\RF\rLWb\LWb):
\eqn\fuse{\hbox{$C^{ijk}\neq 0$ iff $\exists$ roots
$\ai\in\G_i,\aj\in\G_j,\ak\in\G_k$ with $\ai+\aj+\ak=0$.}}
(Note the use here of a convention that will be adhered to for the
rest of this paper: $\ai$, for example, is used for any root that
lies in $G_i$, the orbit of the root $\p_i$. It is important not to
confuse this with the simple root $\a_i$ -- in particular, even
though the labelling of the orbits ultimately derives from the
choice of simple roots, via \thetadf, there are cases where the
simple root $\a_i$ does {\it not} lie in the orbit $\G_i$.)
Since the fusing angles $U^k_{ij}$ are the relative angles of
projections into the $\o^1$ eigenspace, they are related to the
$u(\ai,\aj)$ by
\eqn\urel{U^k_{ij}={\pi\over h}|u(\ai,\aj)|,\quad{\rm where~}
   \ai+\aj\in\G_{\bar k}.}
Being signed angles, the $u(\a,\b)$'s also satisfy
\eqn\ucyc{u(\a,\b)+u(\b,\g)+u(\g,\a)=0\quad {\rm mod~} 2h}
for (any) three roots $\a,\b$ and $\g$.

Armed with these conventions, the expression given in \I\
for the two-particle S-matrix can be rewritten in a compact way:
\eqn\smdfi{S_{ij}=\hprod\ubl{2p+1+u_{ij}}_{\pm}^{(\l_i,w^{-p}\p_j)}.}
Note, the expression is identical in form whether it is written in
terms of $\bl{x}$ or $\mbl{x}$. This is equivalent to
unitarity ($S_{ij}(\t)S_{ij}(-\t)=1$) and follows from
\eqna\phirels
$$\eqalignno{
(\l_i,w^{-p}\p_j)&=-(\l_i,w^{p+1+u_{ij}}\p_j).&\phirels a\cr
\noalign{\hbox{Symmetry ($S_{ij}=S_{ji}$) can also be checked, using}}
(\l_i,w^{-p}\p_j)&=(\l_j,w^{-p-u_{ij}}\p_i).&\phirels b\cr}$$
(A `mixed' equality, $(\l_i,w^{-p}\p_j)=-(\l_j,w^{p+1}\p_i)$,
follows directly from \bourbaki,
independently of the special root ordering \twosplit{}.) Equivalent
formulae were given in \I, but in a less compact way.

Equation \smdfi\ can be put into a perhaps more suggestive form by
noting from \udef\ that $2p+u_{ij}$ is just
$u(\p_i,w^{-p}\p_j)$. Hence
\eqn\smdfib{S_{ij}=
   \prod_{\aj\in\G_j}\ubl{u(\p_i,\aj)+1}_{\pm}^{(\l_i,\aj)}.}
The unitarity and symmetry of this formula can be checked directly
by rewriting \phirels {}\ as
\eqna\alpharels
$$\eqalignno{
(\l_i,\aj)&={-}(\l_i,w^{u(\p_i,\aj)+1}\aj),&\alpharels a\cr
(\l_i,\aj)&=(\l_j,w^{(u(\p_j,\ai)-u(\p_i,\aj))/2}\ai).&\alpharels
b\cr}$$
Note, the exponent of $w$ in \alpharels b\ is an integer,
since $u(\p_j,\ai)$ and $u(\p_i,\aj)$ are always either both even or
both odd. The important case for the symmetry of \smdfib\ is
$(\l_i,\aj)=(\l_j,\ai)$ if $u(\p_j,\ai)=u(\p_i,\aj)$.

The S-matrix bootstrap equations\ts\RF\rZz\Zz\
can be checked very simply from \smdfib.
First these equations are rewritten, for each
particle species $l$ and each nonvanishing
three-point coupling $C^{ijk}$, as
\eqn\si{S_{li}(\t)S_{lj}(\t+iU^k_{ij})S_{lk}(\t-iU^j_{ik})=1.}
(This alternative bootstrap equation, obtained from the
more usual one via the equations of
unitarity ($S_{ij}(\t)S_{ij}(-\t)=1$) and
crossing ($S_{ij}(\t)=S_{i\bar\jmath}(i\pi-\t)$), is analogous
to the symmetrical version of the conserved charge bootstrap equation used
in \I.) The structure of this equation is clarified if a
shift operator $\CT_y$ is introduced, 	defined by
\eqn\ctdef{\bigl(\CT_yf\bigr)(\t)=f(\t+{i\pi y\over h})}
and acting on the blocks as
\eqn\ctprop{\CT_y\ubl{x}_{\pm}=\ubl{x\pm y}_{\pm}.}
Recalling from \fuse\ that $C^{ijk}\neq 0$ implies the existence of
a root triangle $\{\ai,\aj,\ak\}$, the relation
\urel\ can be used to write equation \si\ as
\eqn\sii{\bigl(S_{li}\bigr)\bigl(\CT_{u(\ai,\aj)}
 S_{lj}\bigr)\bigl(\CT_{u(\ai,\ak)}S_{lk}\bigr)=1.}
Note how the fact that $u(\a,\b)$ is a {\it signed} angle takes care
of the relative minus sign between the two shifts in the earlier
equation, \si. Of course, depending on the orientation of the
projection of the root triangle into the $s=1$ subspace, there could
be an overall negation of the shifts in \sii\ compared to \si. In
fact, root triangles projecting to both orientations always exist --
this will be discussed in more detail in section four --
but this is not a problem since \si\ also holds with the labels $j$ and
$k$ exchanged, an operation which itself has the effect of
negating the two shifts.

Finally, acting on both sides with $\CT_{u(\p_l,\ai)}$ and using \ucyc\
gives
\eqn\siii{\(\CT_{u(\p_l,\ai)}S_{li}\)
          \(\CT_{u(\p_l,\aj)}S_{lj}\)
          \(\CT_{u(\p_l,\ak)}S_{lk}\)=1\quad (\ai+\aj+\ak=0).}
To verify that the bootstrap equations in this form
are satisfied by \smdfib\ is
almost immediate. The left hand side of \siii\ involves three roots
running through the orbits $\G_i, \G_j$ and $\G_k$ for the S-matrix
elements $S_{li}, S_{lj}$ and $S_{lk}$ respectively. If these orbits
are labelled starting at $\ai, \aj$ and $\ak$ (instead of the
roots $\p_i, \p_j$ and $\p_k$ that would have been used had \smdfi\
been the starting point), then via \ctprop\ the left hand side is
$$\eqalign{\hprod&
 \ubl{u(\p_l,w^{-p}\ai)\pm
   u(\p_l,\ai)+1}_{\pm}^{(\l_l,w^{-p}\ai)}\cr
\noalign{\vskip-18pt}
\phantom{\hprod}&\qquad\times\ubl{u(\p_l,w^{-p}\aj)\pm
   u(\p_l,\aj)+1}_{\pm}^{(\l_l,w^{-p}\aj)}\cr
\noalign{\vskip-18pt}
\phantom{\hprod}&\qquad\qquad\times\ubl{u(\p_l,w^{-p}\ak)\pm
   u(\p_l,\ak)+1}_{\pm}^{(\l_l,w^{-p}\ak)}.\cr}$$
That this is equal to one is apparent if the
$\mbl{x}$ blocks have been used
($\bl{x}$ being appropriate for the equivalent
version of \siii\ with all shifts negated). From \udef,
for each value of $p$ the three
blocks involved are then equal to the same function,
namely $\mbl{2p+1}$. The total power to which this
is raised is $\(\l_i,w^{-p}(\ai+\aj+\ak)\)$. But this is
zero, since $\ai,\aj$ and $\ak$ form a root triangle for the
coupling $C^{ijk}$. Hence the whole expression is equal to one, as
required.

Section five will give a general proof of
a result for the minimal scattering theories
due to Klassen and Melzer, for which it will be helpful to have an expression
for the minimal S-matrix elements in terms of the unitary blocks
\eqn\uniblock{\usbl{x}=\sbl{x}/\sbl{-x}.}
(This block, unitary in the sense that it individually satisfies the
unitarity constraint mentioned above, was used in \rBCDSc, along with the
larger unitary block $\ubl{x}=\bl{x}/\bl{-x}$.)
Via \block\ and \uniblock, the formulae \smdfi\ and \smdfib\
become, in the minimal cases,
\eqn\smdfii{S_{ij}=\hprod\usbl{2p+u_{ij}}^{(\l_i,w^{-p}\p_j)}
  =\prod_{\aj\in\G_j}\usbl{u(\p_i,\aj)}^{(\l_i,\aj)}.}

\newsec{Pole structure}
This section will involve a detailed  examination of the physical
pole structure of S-matrices described by \smdfi. The
nature of the residues will be studied, for which
a little more information on the building block $\mbl{x}$ will be needed.
This is a $2\pi i$ periodic function, real for imaginary $\t$.
It has simple poles at $(x-1)\pi
i/h$ and $(x+1)\pi i/h$, with residues positive multiples of $-i$, $+i$
respectively. Outside the interval between these two poles (or its repetition
modulo $2\pi i$), the function is positive on the imaginary axis. These facts
hold equally well for the perturbed conformal or the affine Toda blocks.

It is helpful to
introduce a pictorial notation in which the S-matrix element is
represented by a `wall' of rectangles, stacked along the imaginary axis.
The building block $\mbl{x}$ is depicted by a single rectangle
above the
imaginary axis, stretching from $(x-1)\pi i/h$ to $(x+1)\pi i/h$:
\vskip .03in\newdimen\R
\R=2.5in
$$ \mbl{x}~\equiv ~~~\vbox{\offinterlineskip
   \bag{\R}.{\hfil ${-}i$\hfil\hfil ${+}i$\hfil}.
   \vskip4pt
   \bagh{\R}.{\bx10.}.
   \vbox{\hrule\hbox to\R{\hfil\tick0.3.\label{(x-1)\pi i/h}
         \hfil\tick0.3.\hfil\tick0.3.\label{(x+1)\pi i/h}\hfil}}} $$
\vskip .1in
\noindent The residues of the poles at $(x\pm 1)\pi i/h$, up to
some real, positive constant,
are shown above the block. A product of blocks making up an S-matrix element
is represented by stacking the rectangles to make a wall. For example,
$$ \mbl{3}^2\mbl{5}~\equiv ~~~\vbox{\offinterlineskip
\bag{2.7in}.{\hskip.2in\bx10.\nobx}.\bag{2.7in}.{\hskip.2in\bx10.\bx10.}.\vbox
      {\hrule\hbox to2.9in{\hskip.2in
\tick0.3.\label{2\pi i/h}\tline2.0.3.\label{4\pi
i/h}\tline2.0.3.\label{6\pi i/h}\hskip.2in}}} $$
\vskip .08in
\noindent
Poles occur at the ends of the blocks, of higher order where the ends of
blocks coincide. In this example there is a pole of order three at
$\theta =4\pi i/h$, coming
from the two blocks to the left and one to the right.
To represent a full S-matrix element, a wall of length $2\pi$
is sufficient, since $\mbl{x+2h}=\mbl{x}$. In fact, the unitarity
constraint imposes that the height for $\mbl{-x}$ must be exactly the
negative of that for $\mbl{x}$, so a stretch of length $\pi$ will
do; it is convenient to let it straddle the physical strip,
running from $0$ to $i\pi$. To give a couple of examples, here
are two of the S-matrix elements from the $E_8$-related
scattering theories:
\vbox{
\newdimen\H
\H=\hsize
\def \ha{\hskip .5pt}
\def \p#1{\bxes5.#1.} \def \q#1{\nobxes#1.}
\def \nt{\tline9.0.2.\hfil }
\def \etickline{\vbox{\hrule \vskip 0pt
\hbox to \H{\tick0.4.\nt\tick0.3.\nt\tick0.3.\nt\tick0.4.}}}
\def\sclap#1{\clap{$\scriptstyle #1$}}
\def\polemark{\sclap{\B}}
\def\bg#1{\bag\H.#1.}
\def\bgh#1{\bagh\H.#1.}
\def \pece
{{\offinterlineskip \baselineskip =5pt
\bgh{\q4\p2\q2\p2\q4}
\bgh{\p2\q1\p8\q1\p2}
\etickline
\bg{\sclap{0}\nobxh\q2\polemark\q4\polemark\q4\polemark
         \q1\polemark\q3\polemark\nobxh\sclap{i\pi}}
 }}
\def \pehh
{{\offinterlineskip \baselineskip =5pt
\bg{\q5\p5\q5}
\bg{\q4\p7\q4}
\bg{\q3\p9\q3}
\bg{\q2\p{11}\q2}
\bg{\q1\p{13}\q1}
\bg{\p{15}}
\etickline
\bg{\sclap{0}\q{10}\polemark\q1\polemark\q1\polemark\q1\polemark
  \q1\polemark\q1\sclap{i\pi}}
 }}
\def \S#1{\vskip16pt\noindent $S_{#1}$ : \par }
\S{35}
\vskip 8pt
\pece
\S{88}
\vskip 4pt
\pehh
\noindent
}
The rest of each picture, a stretch of wall running from (say) $i\pi$
to $2i\pi$, can be obtained by reflecting the piece shown about the
line Im$(\t)=i\pi$, and then negating all the heights. Note that the
heights are positive inside the physical strip, and negative outside
-- a general phenomenon that will be commented on at the end of this
section.
The positions of expected forward-channel poles (found from the
three-point couplings and the masses) are shown by the symbols
$\B$ below the axis --- they occur precisely at the
`downhill' sections of wall (reading left to right). This apparent
coincidence exemplifies a well-established relationship between
the fusing structure and the nature of the pole residues, and
will now be discussed.

The nature of the
residue of any pole is easy to find from the corresponding
picture\foot{For the affine Toda S-matrices,
this sort of block notation also allows the
value of this residue, to leading order in $\b$, to be
identified; this was used in \rBCDSe.}
(the term residue
being used somewhat loosely to mean the coefficient of the most singular
part). Blocks not contributing directly to a pole simply multiply
its residue by a positive real number, and can be ignored (this is the
reason for the minus sign in \mblock).
Furthermore, the contribution from two directly abutting blocks (one to
the left and one to the right of the pole) is a positive multiple of
$({+}i).({-}i)=1$, so these can also be ignored.
Thus the nature of the residue is determined solely
by the difference in the number of blocks immediately to the left and
right of the pole, that is by the change in height of the wall of
blocks at the position of the pole.
The residues for the three simplest possibilities are
\eqna\res
\R=2.5in
$$ \vbox{\offinterlineskip
   \bag{\R}.{\hfill
\hskip2.4pt\vrule depth10pt\hrulefill}.
   \bag{\R}.{\hrulefill\vrule depth10pt\hskip2pt
\vrule depth10pt\hrulefill}.}
  ~~~\sim~~{-i}~,\eqno\res a$$
$$ \vbox{\offinterlineskip
   \bag{\R}.{\hrulefill\vrule depth10pt\hskip2pt
\vrule depth10pt\hrulefill}.}
  ~~~\sim~~{+1}~,\eqno\res b$$
$$ \vbox{\offinterlineskip
   \bag{\R}.{
\hrulefill\vrule depth10pt\hskip2.4pt\hfill}.
   \bag{\R}.{\hrulefill\vrule depth10pt\hskip2pt
\vrule depth10pt\hrulefill}.}
  ~~~\sim~~{+i}~.\eqno\res c$$
\vskip .15in
\noindent
In cases $a$ and $c$,
the total number of blocks to left and right of the
pole is odd, and so the pole itself is of odd order.
These odd order poles always
have an interpretation in terms of the production
of a bound state. Examination of the ADE scattering theories on a
case-by-case basis has shown that case $c$,
the downhill pole with a $+i$ residue, is always forward
channel, while case $a$, uphill, is crossed channel. The
pictures give a simple `uphill/downhill' mnemonic by which to decide
if an odd-order pole corresponds to a forward
channel bound
state, in agreement with what has already been observed in the two
$E_8$ examples.
(Note, though, that it remains unclear from the point of
view of perturbation theory why the $+i$ residue should be forward
channel for the higher odd-order poles. The mechanism by which
this rule is reproduced, even for the third-order poles,
is quite complicated\ts\rBCDSe.)

Another `empirical' observation can be made:
apart from the
occasional exception at the very edge of the physical strip (for
example, at $0$ and $i\pi$ in $S_{88}$ above)
the wall height never
changes by more than $\pm 1$. In the exceptional cases, the height
change is always from $-1$ to $1$ or back.
Wall segments of negative height have
zeroes instead of poles, so there is a cancellation and
S-matrix is analytic at
these points (and in fact is, by unitarity, forced to be equal to
$\pm 1$).
Hence \res {}\
turns out to cover all possibilities for S-matrix poles. In
particular, even-order poles always have positive real
residues.

It might be expected that all
the observations described above should have a universal explanation
in the context of root systems. In fact, this can be achieved
using only the most elementary properties of the simply-laced
roots.

Referring back to \smdfib, consider
the pole in $S_{ij}$ at relative rapidity ${\pi i\over h}u(\p_i,\aj)$
(with $0<u(\p_i,\aj)<h$ for the physical strip).
The two blocks contributing to this
pole involve $\aj$ and $w\aj$, that involving $w\aj$ being to the left
(recall from \udef\ that $u(\p_i,w\aj)=u(\p_i,\aj)-2$). Thus the
change $\d h$ in wall height at this pole is given by
\eqn\hdif{ \d h = (\l_i,\aj)-(\l_i,w\aj).}
Using \bourbaki\ for the second equality, this simplifies:
\eqn\hdiff{\d h=\((1-w^{-1})\l_i,\aj\)=(\p_i,\aj).}
Being the inner product of two roots (of a simply-laced algebra),
it is now clear that the change in wall height, if not zero, can only be
$\pm 1$ or $\pm 2$. These possibilities can be examined in turn.

A change of ${-}1$ should correspond to a bound state. But
$(\p_i,\aj)=-1$ implies that $\p_i+\aj$ is a root, $-\ak$ say, since
it is just the Weyl reflection of $\p_i$ with respect to $\aj$.
This gives a root triangle $\{\p_i,\aj,\ak\}$ and
from the fusing rule a non-zero three-point coupling $C^{ijk}$.
The fusing angle for the $\bar k$ bound state
is, by \urel, ${\pi\over h}u(\p_i,\aj)$ --
exactly that corresponding to the pole under discussion.
 Conversely if there {\it is} a
three-point coupling such that a bound state at the relevant fusing
angle is a possibility, then $\d h=-1$ is forced (since
$\p_i+\aj=-\ak$ is then a root, and so $\d h=
(\p_i,\aj)={1\over 2}(\ak^2-\p_i^2-\aj^2)=-1$).

The story for $\d h=1$ is similar, with the conclusion that $\d
h=1$ if and only if there is a bound state in the crossed channel.

These two cases have taken up both forward and crossed
channels, so an even-order pole can {\it never} have an
associated single-particle bound state. It only remains to remark
that $\d h=\pm 2$ implies $\p_i=\pm\aj$, and a relative rapidity for
the putative pole of $0$ or $i\pi$, the first to be found in
$ii$ scattering, the second in $i\bar\imath$. As already mentioned,
such `poles' disappear by unitarity.

This then provides a universal explanation for the interplay between
the bootstrap equations and the pole structure of the purely elastic
S-matrices. In particular it elucidates the `internal' consistency
of these S-matrices, obeying as they do the very equations that they
imply via their pole structure.

To close this section, a comment on a slightly simpler feature of
\smdfi\ and \smdfib, which also has an interpretation in terms of
root systems. Over the full range from $0$ to $2\pi i$,
the wall height can be both positive and negative,
and indeed must be so to satisfy the unitarity constraint.
But on physical grounds, the height had better not be negative for
blocks in the physical strip: in such an eventuality, a perturbed
conformal theory S-matrix would have zeroes in the physical strip,
while the affine Toda theory would gain physical-strip poles with
coupling-constant dependent positions. Now this height is given as
the inner product of some root with a fundamental weight, and so is
positive or negative according to whether the relevant root is
positive or negative with respect to the given set of simple roots
(to say this in another way,
the wall height around ${\pi i\over h}\(u(\p_i,\aj)+1\)$
is just one `component' of the height of the root $\aj$, namely that
piece due to the simple root $\a_i$).
Referring in particular back to \smdfi, it is clear that the physical
requirement reduces, traversing each orbit $\G_j$ of the Coxeter
element starting at the special root $\p_j$,
to a discussion of which roots are positive and which negative.
Although the details will be omitted here, it is straightforward to
use results (i), (ii) and (iii) from the beginning of section two,
together with the implication from unitarity that approximately
half\foot{in fact
exactly half, except for $A_{2n}$ for which $h$ is odd and the
situation is marginally more complicated.} of the roots in a given
orbit are positive, to see that the desired property of the wall
heights does indeed follow from general theory.

\newsec{The signs of the three-point couplings}
In the perturbative treatment of the affine Toda theories, expansion
of the potential to order $\b$ results in a set of three-point
couplings $C^{ijk}$ which, if non-vanishing, obey the `area rule':
$$C^{ijk}=\sg^{ijk}{4\b\over\surd h}\Delta_{ijk},$$
where $\Delta_{ijk}$ is the area of a triangle of sides $m_i,m_j$
and $m_k$ (the particle masses) and $\sg^{ijk}$ is a phase of
unit modulus, which given the hermiticity of the original Lagrangian
can be taken to be plus or minus one.
The vanishing/non-vanishing of the coupling is
described by the rule \fuse. This, together with the normalisation
of $C^{ijk}$, has now been derived in a general
way\ts\RF\rFLOa{\Fb\semi\FLOa}.
However the signs $\sg^{ijk}$ are a little more subtle. Clearly
they can be changed around by negating some of the fields, but this does
{\it not} mean that they can all be set to $1$. Indeed,
cancellations necessary for perturbation theory to be compatible with
integrability often depend crucially on the presence of relative phases
between different terms, which would not be present if all the
$\sg$'s were equal to $1$ (for some examples of this, see
\NRF\rBSa\BSa\refs{\rBCDSe,\rBSa}).

Despite the apparent arbitrariness involved, there is a special
choice of normalisations which connects with the root system data
already described.
Recall that any three-point coupling $C^{ijk}$ results in a
bound-state pole in $ij$ scattering of odd order $2m+1$,
and note that the field normalisations used in \rBCDSc\ can be
altered so that in every case,
\eqn\signrule{\sg^{ijk}=(-1)^m.}
This is actually only a small increase in information over a
formula given in \rBCDSe, itself a consequence of a formula found by
Braden and Sasaki\ts\rBSa. Their result reads:
\eqn\harry{\sg^{ijk}=-\sg^{il\bar m}\sg^{jm\bar n}\sg^{kn\bar l},}
holding in this form whenever the triangle $\Delta_{ijk}$ is tiled
internally by the three other mass/coupling triangles, $\Delta_{il\bar
m},\Delta_{jm\bar n}$ and $\Delta_{kn\bar l}$. Note, \harry\ does
not change with changes to the field normalisations.
The consequence of this, remarked in \rBCDSe, is
that if $\Delta_{ijk}$ is tiled in a `nested' fashion by $2q+1$
other triangles $\{\Delta_A\}$, $A=1,\dots 2q+1$ (with phase
factors $\{\sg^A\}$), then
\eqn\prodrule{\sg^{ijk}=({-}1)^q\prod_A \sg^A.}
(The tiling of a triangle is nested
if it is tiled by three other triangles each of
which is either untiled, or is itself tiled in a nested way.
 This allows \signrule\ to be used
inductively to derive \prodrule.)

Now higher poles are also associated with tilings by mass
triangles\rBCDSe. For an odd-order pole of order $2m+1$,
with $ij$ producing a bound state $\bar k$, the triangle
$\Delta_{ijk}$ has a `maximal' nested tiling (in fact, many such) by $2m+1$
triangles, maximal in the sense that each constituent triangle cannot be
further tiled. The number $2m+1$ was called the depth of the coupling
triangle $\Delta^{ijk}$ in \rBCDSe. Applying \prodrule\ then gives the
phases for all coupling triangles as products of the phases for
triangles of depth 1. It
is then only necessary to check that all the unit depth
phases can all be set to
1 to deduce \signrule\ from \prodrule.

Now \signrule, together with the discussion of the last section,
can be used to give an
expression for the signs in terms of the roots and weights.
The physical strip pole for the $\bar k$ bound state in $ij$
scattering will be at ${\pi i\over h}u(\p_i,\ajo)$ for some $\ajo\in\G_j$
({\it cf} the discussion preceding \hdif). Note,
$\{\p_i,\ajo,\ako\}$, the corresponding root triangle for
$C^{ijk}$, is oriented such that $0<u(\p_i,\ajo)<h$.
The order of the pole is
\eqn\hsum{ 2m+1 = (\l_i,\ajo)+(\l_i,w\ajo).}
Now the change $\delta
h$ in the wall height at this point is $-1$, as the pole is forward
channel; so combining \hsum\ with \hdif\ gives
\eqn\mpd{m=(\l_i,\ajo)}
and $\sg^{ijk}=({-}1)^{(\l_i,\ajo)}$.
This expression for $\sg^{ijk}$ involves the choice of one particular
root triangle,
and furthermore its symmetry in $i,j$ and $k$ is not at all obvious.
To remedy these defects, the idea of the orientation of a root
triangle will have to be made a little more precise. Let
$\{\ai,\aj,\ak\}$ be any root triangle implying the
nonvanishing of the coupling $C^{ijk}$; the orientation
$\epsilon(\ai,\aj,\ak)$ is then defined to be ${+}1$ if the projection
into the $s=1$ subspace has a clockwise sense (going from $i$ to $j$
to $k$), and ${-}1$ if anticlockwise. Since
$\epsilon(\ai,\aj,\ak)={+}1$ if and only if
$0<u(\ai,\aj)<h$, the root triangle
$\{\p_i,\ajo,\ako\}$ used above has orientation ${+}1$.
It will be helpful to define another quantity for (general) root
triangles, namely
\eqn\fdef{f(\ai,\aj,\ak)={1\over 2}+(R\ai,\aj),}
where $R$, given by \rdef, has the important property that
$R\p_i=\l_i$. Finally, set
\eqn\mef{m(\ai,\aj,\ak)
   =\epsilon(\ai,\aj,\ak)f(\ai,\aj,\ak)-{1\over 2}.}
For the original root triangle $\{\p_i,\ajo,\ako\}$, this coincides with
\mpd\ (this is the reason for the ${-}{1\over 2}$).
But $m(\ai,\aj,\ak)$
is the same for {\it any}
triangle of roots for
$C^{ijk}$, and furthermore is symmetrical in $i,j$ and $k$.
To establish these properties, some more information on the set of root
triangles is needed.
There are in fact $2h$ ordered triplets $\{\ai,\aj,\ak\}$
for each non-zero three-point coupling (the ordering being needed to
count correctly the cases when, say, $i=j$).
Of these, $h$ can be found simply by acting `diagonally'
with powers of
$w$ on an initial triangle. These all have the same orientation.
To see that there are exactly $h$ more,
consider the action of
$\tw\equiv -\wb$ (one could equally well set $\tw=\ww$;
all the identities to be given below would be unchanged).
As mentioned in \I, on the cosets $\tw$ has the effect of two successive
charge conjugations, that is no effect at all: $\tw\G_i=\G_i$.
But when acting on triangles, $\tw$ reverses the orientation:
\eqn\otw{\epsilon(\tw\ai,\tw\aj,\tw\ak)=-\epsilon(\ai,\aj,\ak).}
Applying powers of $w$ to $\{\tw\ai,\tw\aj,\tw\ak\}$ then
gives $h$ more possibilities, and the geometry of the projection into the
$s=1$ subspace shows that there can be no
more. (To be more precise, since $U^k_{ij}$ is fixed,
\urel\ shows that once the root $\ai$ has been chosen
there are only two possible directions for the $s=1$ projection of
$\aj$.
But the roots in a single orbit all project to different directions
for $s=1$, so there are at most two roots in $\G_j$ which can form a
triangle involving $\ai$ and a root from $\G_k$. Letting $\ai$ run
through its orbit then gives a maximum of $2h$ triangles.)

An important consequence of the above is that
all triangles for a given coupling
can be obtained from an initial one by acting with
$w$ and $\tw$. Now the effect of these two operations on the
orientation has already been given: $w$ leaves $\epsilon$ invariant,
while $\tw$ negates it. It is also clear that $w$ leaves $f$
unchanged, $w$ being an orthogonal transformation which commutes with
$R$. The action of $\tw$ is a little harder to see, but
the identity $\tw R\tw=1-R$ together with the fact that the inner
product of $\ai$ and $\aj$ must be ${-}1$ (their sum being another
root) implies
\eqn\ftw{f(\tw\ai,\tw\aj,\tw\ak)=-f(\ai,\aj,\ak).}
Comparing with \otw\ establishes the invariance of $m(\ai,\aj,\ak)$
under the diagonal action of both $w$ and $\tw$, and hence its
insensitivity to the choice of root triangle.

There remains the symmetry of \mef\ between $i,j$ and $k$. The
orientation is completely antisymmetric in its arguments, so it will
be enough to demonstrate that the same is true of $f$.
First consider swapping $\ai$ and $\aj$ in \fdef. From \rpropi, and
using again that the inner product of $\ai$ and $\aj$ must be
${-}1$, this sends $f$ to ${-}f$. Now consider a cyclic permutation of
$\ai,\aj$ and $\ak$. The three roots sum to zero, so
$(R\aj,\ak)=(R\aj,-\ai-\aj)=(R\ai,\aj)$, using \rpropi\
and also the fact that $(R\aj,\aj)=1$. Thus $f$ is unchanged by a cyclic
permutation of its arguments, and this is enough to establish
antisymmetry.

This completes the proof of the claims following equation \mef.
The expression can now be used in \signrule, it already having
been mentioned that for one particular triangle (and hence for all)
\mef\ coincides with the previously-derived \mpd.
This can be used in a complete specification of the three-point couplings
for the simply-laced affine Toda theories.
With a simple rewriting of the resulting expression for $({-}1)^m$,
the three-point coupling data can be summarised as follows:

\item{$\B$}$C^{ijk}\neq 0 $ iff $\exists\ \ai+\aj+\ak=0$
(where $\ai\in\G_i,\aj\in\G_j,\ak\in\G_k$).
\item{$\B$} There is a normalisation of the fields
such that in these cases
$$C^{ijk}=
\epsilon(\ai,\aj,\ak)(-1)^{(R\ai,\aj)}{4\b\over\surd h}\Delta_{ijk}.$$

\newsec{A formula due to Klassen and Melzer}
In their investigations of the Thermodynamic Bethe Ansatz, Klassen and
Melzer\ts\RF\rKMa\KMa\ observed an interesting universal feature of the
minimal purely elastic S-matrices. If the matrix $N_{ij}$ is defined by
\eqn\ndef{N_{ij}=-{1\over 2\pi i}\Bigl[ \ln
S_{ij}(\t)\Bigr]_{\t=-\infty}^{\t=\infty}}
then for the minimal (perturbed conformal field theory) S-matrices,
\eqn\kmresult{N=2C^{-1}-I,}
where $C$ is the relevant (non-affine) Cartan matrix, and $I$ the
unit matrix. Using this
result, Klassen and Melzer were able to calculate the central charges
of the ultra-violet limits of these theories, finding agreement with
the idea that these S-matrices do indeed describe perturbations of
certain conformal field theories. Such issues will not be the concern
here; the purpose of this section is merely to give a simple and
universal proof of \kmresult\ starting from the general S-matrix
expression of \I.

Klassen and Melzer used unitary blocks $f_{\a}(\t)\equiv\usbl{h\a}(\t)$,
in terms of which the S-matrix element $S_{ij}$ was written
$$S_{ij}=\prod_{\a\in A_{ij}}f_{\a}(\t).$$
The parameter $\a$ was taken to satisfy $-1 < \a\leq 1$,
and \ndef\ reduced to
\eqn\kndef{N_{ij}=\sum_{\a\in A_{ij}}(1-|\a|){\rm sgn}(\a),}
with the convention that ${\rm sgn}(0)=0$ (since $f_0\equiv 1$).
It will be more convenient below to
choose a different range for $\a$, namely
$0\leq\a <2$. Using $f_{\a}=f_{\a+2}$ to
relabel the negatively-indexed blocks,
together with the fact that
$(1-|\a|){\rm sgn}(\a)=1-(\a+2)$ for $-1<\a<0$, equation
\kndef\ becomes
\eqn\kdef{N_{ij}=\sum_{\a\in A_{ij}}(1-\a)-|A_{ij}\cap\{0\}|,}
where now $0\leq\a<2$, and the second term
undoes the overcounting in the first whenever $\a=0$.

To evaluate \kdef, it is simplest to use the first expression of
\smdfii. Each unitary block $\usbl{2p+u_{ij}}$
corresponds to an $\a\in A_{ij}$ equal to
$(2p+u_{ij})/h$,
while $|A_{ij}\cap\{0\}|$ is
simply the number of blocks $\usbl{0}$.
Such blocks will only be found if $u_{ij}=0$, and so
$|A_{ij}\cap\{0\}|=0$ if $i$ and $j$ have different colours. Otherwise,
$\usbl{0}$ is raised to the power
$(\l_i,\p_j)$. For the
the ordering \twosplit{}\ of the simple roots, the full set of these
inner products is
\eqnn\dotprop
$$\vcenter{\openup1\jot \tabskip=0pt plus1fil
\halign {\tabskip=0pt
$#\hfil$&${}#\hfil$\cr
(\l_{\B},\p_{\B'})=\delta_{\B\B'}&~~~~(\l_{\W},\p_{\B'})=-C_{\W\B'}\cr
(\l_{\B},\p_{\W'})=0&~~~~(\l_{\W},\p_{\W'})=\delta_{\W\W'}\cr}}
\eqno\dotprop$$
Hence the general result is $|A_{ij}\cap\{0\}|=\d_{ij}$,
and $N_{ij}$ is given by
\eqn\myndef{N_{ij}=
\sum_{p=0}^{h-1}(1-(2p+u_{ij})/h)(\l_i,w^{-p}\p_j)-\d_{ij}.}
Strictly speaking, the fact that $u_{\B\W}=-1$ means that
if $i$ is of type $\B$ and $j$ of type $\W$,
the first block counted by \myndef\ corresponds to $\a={-}1/h$, outside the
desired range. This can be ignored as the power to which this block
is raised, $(\l_i,\p_j)$, is zero here by \dotprop.

Making use of equations \rprop\ and \projop,
$$N_{ij}=\bigl( \l_i,\bigl( (h-u_{ij})P_0+2R\bigr)\p_j\bigr)-\d_{ij}
=2(\l_i,\l_j)-\d_{ij},$$
the results $P_0=0$
($0$ is never an exponent) and $R\p_j=\l_j$ giving
the second equality.
Since $(\l_i,\l_j)$ is exactly the inverse of the
Cartan matrix, equation \kmresult\ now follows immediately.

\newsec{Conclusions}
Various previously-observed
features of the ADE purely elastic scattering theories are now
known to follow from general principles,
especially given recent work on affine Toda perturbation
theory\ts\rFLOa\ and the Clebsch-Gordan rule for fusings\ts\RF\rBa\Ba. With
regards to the S-matrices, a
notable feature of the treatment given above, as compared to that
in \I, is that the splitting of the roots according to type
(colour) has become much less important. Essentially, once the
notations \udef\ have been set up, this distinction can
be forgotten. All that is needed is equation \urel, relating the
quantities $u(\a,\b)$ to the fusing angles.
However, it should not be forgotten that
the splitting of the particles into two sets {\it does} express a geometrical
property of the projections of their orbits, reflected in the
fact\ts\I\ that two particles of the same type always fuse at an even
multiple of $\pi/h$, two particles of opposite type at an odd multiple of
$\pi/h$. Hence the split still has physical implications, even if these are
best left hidden for most calculations.

As regards future work, there are two obvious questions to ask.
Within the context of theories with diagonal S-matrices, are there
any other physically reasonable possibilities beyond those
associated with the ADE series? To
formulate this question properly, it must be decided exactly what is
meant by `physically reasonable'. For example, whenever a subalgebra
of the fusing algebra can be formed (such as emerges when a twisted
folding is considered\ts\rBCDSc\rCMa), a self-consistent set of S-matrix
elements, obeying the bootstrap equations that they imply via their
odd-order poles, can be obtained simply by taking the corresponding
submatrix of the `parent' S-matrix. However, in such cases there are always
higher-order physical poles which are inexplicable without the full
set of particles of the parent theory, forcing their re-inclusion
and returning the theory to the ADE set.
(A similar phenomenon allows the solitons of the sine-Gordon theory
to be inferred from the S-matrix elements of the breathers
alone\ts\RF\rGa\Ga.) The only other purely elastic S-matrices that seem
to have been discussed so
far\ts\NRF\rCMc\CMc\NRF\rFKMa\FKMa\refs{\rCMc,\rFKMa}\
all obey rather
different self-consistency requirements, in that
the prescription for assigning forward and crossed
channels to the odd-order poles is changed.
Such conditions appear to be appropriate for theories with non-hermitian
lagrangians\ts\RF\rCMc\CMc, and so these S-matrices
should in any case fall outside an
initial (perhaps ADE)
classification of unitary purely elastic scattering theories.
Nevertheless, it would be interesting if they could also
be given an interpretation in terms of root systems.

The second question concerns the relevance of any of the above to
theories with multiplets and non-diagonal S-matrices. This would
require some similarity in structure between the bootstraps for the purely
elastic scattering theories and those for at least a subset of the
more complicated models. At least at the level of
the fusings, the same structure  has emerged in the study of perturbations
of $N{=}2$ supersymmetric conformal theories\ts\rLWb\ (although in fact not
for all the ADE series). Other possible
candidates for inclusion in the
subset include the principal chiral models\ts\RF\rOWa\OWa. S-matrices
with the same scalar (CDD) part have been proposed for perturbations of
certain conformal field theories\ts\RF\rBLa\BLa. There are many
coincidences (for example, of mass spectra) and even some explicit
calculations\ts\RF\rMd\Md\ suggesting that a connection
may indeed be found,
but the greatly-increased complexity of the bootstrap equations
makes a complete analysis difficult.

\bigskip\centerline{\bf Acknowledgements}
\smallskip
I would like to thank Ed Corrigan for interesting discussions and
comments on the manuscript, and the Research Institute for
Mathematical Sciences, Kyoto University for their kind hospitality while this
work was being completed. The work was supported in part by a grant
of the British Council. I am grateful to the Royal Society for a
Fellowship under the European Science Exchange Programme.

\listrefs
\end